\documentclass[preprint,authoryear,12pt]{elsarticle}

\usepackage{amssymb}
\usepackage{amsmath}
\usepackage{amsthm}
\usepackage{epstopdf}
\usepackage{setspace}
\usepackage{rotating}
\usepackage[T1]{fontenc}
\usepackage{subfig}
\usepackage{algorithm2e}
\usepackage{chngcntr} 
\usepackage{lineno}
\usepackage[framed,numbered,autolinebreaks,useliterate]{mcode}
\usepackage{verbatim} 

\newtheorem*{theorem*}{Theorem} 

\usepackage{epstopdf}
\usepackage{adjustbox}
\usepackage{blindtext}
\journal{arXiv.org}
\begin{document}
\begin{frontmatter}
\title{Whole genome single nucleotide polymorphism genotyping of \textit{Staphylococcus aureus}}
\author{Changchuan Yin$^{1,\ast}$, Stephen S. -T. Yau$^{2,\ast}$}
\address{1. Department of Mathematics, Statistics, and Computer Science\\ The University of Illinois at Chicago, Chicago, IL 60607-7045, USA\\

2. Department of Mathematical Sciences\\ Tsinghua University, Beijing 10080, China\\
$^{\ast}$Corresponding authors, cyin1@uic.edu, yau@uic.edu
}
\begin{abstract}
Next-generation sequencing technology enables routine detection of bacterial pathogens for clinical diagnostics and genetic research. Whole genome sequencing has been of importance in the epidemiologic analysis of bacterial pathogens. However, few whole genome sequencing-based genotyping pipelines are available for practical applications. Here, we present the whole genome sequencing-based single nucleotide polymorphism (SNP) genotyping method and apply to the evolutionary analysis of methicillin-resistant \textit{Staphylococcus aureus}. The SNP genotyping method calls genome variants using next-generation sequencing reads of whole genomes and calculates the pair-wise Jaccard distances of the genome variants. The method may reveal the high-resolution whole genome SNP profiles and the structural variants of different isolates of methicillin-resistant \textit{S. aureus} (MRSA) and methicillin-susceptible \textit{S. aureus} (MSSA) strains. The phylogenetic analysis of whole genomes and particular regions may monitor and track the evolution and the transmission dynamic of bacterial pathogens. The computer programs of the whole genome sequencing-based SNP genotyping method are available to the public at https://github.com/cyinbox/NGS.

\end{abstract}

\end{frontmatter}
 
 \section{Introduction}
 \label{Introduction}
 \textit{Staphylococcus aureus} (\textit{S. aureus}) is a type of bacteria that some people carry in their noses. This organism typically does not cause any health problems; however, it sometimes causes hospital and community acquired infections. \textit{S. aureus} can cause serious infections such as bloodstream infections, pneumonia, or bone and joint infections. The epidemiology of \textit{S. aureus} was a series of waves of resistance ~\citep{chambers2009waves}. Methicillin-resistant \textit{S. aureus} (MRSA) generates growing concerns over its resistance to almost all antibiotics except vancomycin. MRSA infections are associated with significant morbidity and mortality compared to infection due to methicillin-susceptible \textit{S. aureus} (MSSA). MRSA strains carry either a mecA- or a mecC-mediated mechanism of resistance to a broad range of beta-lactam antibiotics. The MRSA transmission poses a challenge to infection prevention and public health. The evolving spread of MRSA emphasizes the importance of surveillance, infection control vigilance and the ongoing investigation of rapid typing methods for MRSA.
 
 The molecular typing technique, pulsed-field gel electrophoresis, is routinely used to study the epidemiology and outbreak of \textit{S. aureus} because of its high discrimination power ~\citep{deurenberg2008evolution,comas2009genotyping}. However, various experimental conditions and subjective decisions about DNA banding patterns may result in an inconsistent classification of \textit{S. aureus} clones. The multilocus sequence typing (MLST) involves sequencing DNA fragments of seven housekeeping genes. The sequences of these genes are compared to known alleles at each locus via the MLST database, where every isolate is described by a seven-integer allelic profile that defines a sequence type (ST). MLST is typically preferred as a general typing technique because the data can be exchanged between different laboratories. However, MLST is expensive and time-consuming ~\citep{deurenberg2008evolution}. In addition, since MLST based genotyping only involves specific core regions, mutations in the whole genome are unidentified. The genome sequence genotyping for bacteria is by Variable Number Tandem Repeats (VNTR) ~\citep{comas2009genotyping}. However, VNTR-Based genotyping results in unreliable phylogenetic inference ~\citep{comas2009genotyping}.  
 
 Single nucleotide polymorphism (SNP) screening has been used in analyzing and associating traits with regions of the genome for human genomes ~\citep{international2001map,yu2017single}. As sequencing costs continue to decline, next-generation sequencing becomes a novel approach for genotyping of organisms. The whole genome sequencing based SNP genotyping can represent an effective method for characterizing and profiling bacterial strains ~\citep{schurch2018whole}, and therefore may play a critical role in the surveillance of MRSA outbreak and transmission. Bacterial SNP genotyping can be performed via mapping raw reads to reference genomes, without assembled genomes. This enables a fast screening and investigation of bacterial pathogens during emergency outbreaks. The currently available microbial SNP pipelines are Snippy~\citep{seemann2015snippy} and SNVphyl~\citep{petkau2017snvphyl}. Snippy finds SNPs between a haploid reference genome and the Next-Generation Sequencing reads and follows alignment for phylogeny construction. However, Snippy is time-consuming and may not construct phylogeny for large genomes because of the nature of multiple alignments. 
 
 The SNP genotyping method gives sufficient resolution of genome mutations and structural variants and can be considered as the genome fingerprint. The whole genome sequencing-based SNP genotyping may reveal larger recombination events in \textit{Streptococcus pneumoniae} ~\citep{roe2016whole,cowley2018evolution}. Despite being used in many core gene analyses, few of the whole genome sequencing-based SNP genotyping method is available to general research communities and hence is not widely used in the phylogenetic analysis of bacterial pathogens, especially in MRSA genome analysis. 
 
 In this study, we present the whole-genome sequencing-based SNP genotyping method to investigate the epidemiology of global MRSA. The phylogenetic analysis of MRSA and MSSA strains demonstrates that the whole genome sequencing-based SNP genotyping may differ various isolates and show the epidemiology dynamics from conversed regions and high variant regions in genomes. As demonstrated, the proposed SNP genotyping method can be effective and practical to explore the epidemiology of MRSA and other microorganisms.
 
 \section{Methods}
 \subsection{Whole genome sequencing-based SNP genotyping}
 The whole genome sequencing-based genotyping is to associate the whole genome variants with the traits of an organism. The genome variants include single nucleotide polymorphisms (SNPs), structural insertions, and deletions. We implement the following pipeline to perform the variant calling and SNP genotyping from whole genome sequencing reads.
 
 The whole genome paired-end DNA sequencing reads were mapped on the reference genome \textit{S. aureus} NCTC8325 ~\citep{gillaspy2006staphylococcus}. This reference strain was originally used as a propagating strain for bacteriophage 47 and is considered the prototypical strain for most genetic research on \textit{S. aureus}. The reference genome was first indexed by aligner bwa ~\citep{li2009fast}. The raw reads were aligned to the indexed reference genome using bwa Burrows-Wheeler transform ~\citep{li2009fast,ta2018workflow}. The aligned sam file was converted and sorted to bam file by samtool ~\citep{li2009sequence}, and then indexed by picard ~\citep{picard2018}. The aligned reads were called for variants by the Genome Analysis Toolkit ~\citep{mckenna2010genome}. The coverage analysis of the read mapping was performed using bedtools ~\citep{quinlan2010bedtools}. The resulted variant calling file was analyzed by custom Python programs, which use scikit-alle library to process the variant calling files ~\citep{scikit-allel} and are available at https://github.com/cyinbox/NGS. The sliding window is used to calculate the histogram of SNPs along the genome variants. The pipeline is described as Algorithm 1.
 \begin{algorithm} 
 	\SetAlgoLined
 	\KwIn{a reference genome (fasta), pair-ends reads (fastq)}
 	\KwOut{variant calling (vcf file)}
 	\textbf{Step:}
 	\begin{enumerate}
 		\item Index the reference genome (bwa).
 		\item Map paired-end sequences to the reference genome (bwa).
 		\item Index the reference genome again for faindex (samtools).
 		\item Convert the sam file to a bam file for fast processing speed (samtools).
 		\item Sort the bam file (samtools).
 		\item Index the sorted bam file (samtools).
 		\item Create the dictionary of the contig names and sizes of reference genome (picard).
 		\item Add read groups into the sorted bam file (picard).
 		\item Build index on the sorted bam file (picard).
 		\item Variant calling (GATK).
 		\item Perform read mapping coverage (bedtool).
 	\end{enumerate}
 	\caption{The pipeline for variant calling by whole genome sequencing-reads on a reference genome. The software used in the pipeline are included in the steps.}
 \end{algorithm}
 
 \subsection{Distance of SNP variants}
 To compare the difference between the SNP variant profiles of genomes, we use the Jaccard distance as a measure. The Jaccard similarity coefficient is defined as the intersection size divided by the union size of two sets $A, B$ ~\citep{levandowsky1971distance}.
 \begin{equation*}
 J(A,B) = \frac{{\left| {A \cap B} \right|}}
 {{\left| A \right| + \left| B \right| - \left| {A \cap B} \right|}}
 \end{equation*}
 
 The Jaccard distance is a metric on the collection of all finite sets. The Jaccard distance of two sets $A$ and $B$ is scored by one subtracting the Jaccard similarity coefficient. 
 \begin{equation*}
 d_J (A,B) = 1 - J(A,B) = \frac{{\left| {A \cup B} \right| - \left| {A \cap B} \right|}}
 {{\left| {A \cup B} \right|}}
 \end{equation*}
 
 The SNP set comparison in the Jaccard distance measure is in the corresponding positions of genomes, therefore, the Jaccard distance also takes account of the ordering of SNPs. The genetic distance of two bacterial strains is measured as the Jaccard distance of SNP variants of the bacterial genomes ~\citep{comas2009genotyping,yu2017novel}.
 
 \subsection{Whole genome sequencing-based phylogenetic analysis}
 We select and concatenate SNP rare or rich regions of the genomes according to the histograms of SNPs. The threshold for selecting the SNP rare or rich regions is 10 SNPs in a sliding window of length 250 bp. The phylogenetic tree of different genomes can be constructed based on the Jaccard distances of these genomic segments or whole genomes. In this study, the UPGMA (Unweighted Pair Group Method with Arithmetic Mean) phylogenetic trees, a bottom-up hierarchical clustering method, are built based on pair-wise Jaccard distances of SNP profiles in genomes. The algorithm for the phylogenetic analysis is outlined as Algorithm 2. 
 
 \begin{algorithm} 
 	\SetAlgoLined
 	\KwIn{variant callings of genomes}
 	\KwOut{phylogeneic tree}
 	\textbf{Step:}
 	\begin{enumerate}
 		\item Compute histogram of SNPs each sliding window.
 		\item Concatenate the genomic fragments that have at least 10 SNPs per sliding window. 
 		\item Concatenate the genomic fragments that have less than 10 SNPs per sliding window. 
 		\item Compute pair-wise Jaccard distances of high SNPs or low SNPs regions.
 		\item Construct the UPGMA phylogenetic trees using the Jaccard distances.
 	\end{enumerate}
 	\caption{Whole genome sequencing-based phylogenetic analysis.}
 \end{algorithm}
 \subsection{DNA sequencing data}
 The complete reference genomes were retrieved from the National Center for Biotechnology Information (NCBI) GenBank repository ~\citep{benson2008genbank}. Whole genome sequencing paired-end raw reads were downloaded from NCBI Sequence Read Archive (SRA) as Fastq format ~\citep{kodama2011sequence}. The sequence reads of the genomes were originally generated by Illumina HiSeq series instruments by various research groups. The reference genomes and SRA sequence reads are listed in Table 1. It should be noted that SNP genotyping uses paired-end DNA sequencing reads, which provides superior alignment across repetitive regions and also detects rearrangements such as insertions, deletions and inversions ~\citep{korbel2007paired}.
 
 \begin{table}[ht]
 	\caption{The whole genome sequencing reads of S. aureus genomes.}
 	\centering 
 	\begin{tabular}{l*{5}{l}r}
 		\hline\hline
 		GenBank or SRA ID  & strain name & MRSA &  \\
 		\hline
 		NC007795 & \textit{S. aureus} NCTC8325&-&\\
 		BX571856  & \textit{S. aureus} ST252&+&\\
 		SRR6475664  & \textit{S. aureus} ST398&+&\\
 		SRR6304955 & \textit{S. aureus} HPV107&+&\\
 		SRR7295360 & \textit{S. aureus} HPV107&+&\\
 		SRR7295358 & \textit{S. aureus} NRS2&+&\\
 		ERR2541869 & \textit{S. aureus} ST88&+&\\
 		ERR2275539 & \textit{S. aureus} MSSA-HWC2014&-&\\
 		SRR5714648 & \textit{S. aureus} MSSA-H41&-&\\
 		SRR6675978 & \textit{S. aureus} MSSA-ST97&-&\\
 		ERR2541870 & \textit{S. aureus} Nigeria&+&\\
 		ERR377327  & \textit{S. aureus} USA300 RU-CAMP-29c&+&\\
 		ERR2276455 & \textit{S. aureus} USA300-RU-CAMP-P29a&+&\\
 		ERR2276459 & \textit{S. aureus} USA300-RU-CAMP-P29b&+&\\
 		ERR2541868 & \textit{S. aureus} ST88-a&+&\\
 		ERR2541871 & \textit{S. aureus} ST88-b&+&\\
 		ERR2562460 & \textit{S. aureus} CC398-ST899&+&\\
 		ERR1795563 & \textit{S. aureus} MSSA-SHAIPI&-&\\
 		SRR6304955 & \textit{S. aureus} PtA02-T1&+&\\
 		SRR6304957 & \textit{S. aureus} PtA04-T3&+&\\
 		SRR1014694 & \textit{S. aureus} USA300 OR-54&+&\\
 		SRR5617496 & \textit{S. aureus} USA300 22862\_R1&+&\\
 		SRR1014700 & \textit{S. aureus} USA500-NRS385E&+&\\
 		SRR1014703 & \textit{S. aureus} USA400-BAA1752 CC1-ST1&+&\\
 		SRR1014705 & \textit{S. aureus} CC80-24329-ST153&+&\\
 		SRR1014706 & \textit{S. aureus} USA700-NRS386 CC72-ST72&+&\\
 		SRR1014707 & \textit{S. aureus} USA700-GA-442 CC72-ST72&+&\\
 		SRR1014708 & \textit{S. aureus} USA800-NRS387 CC5-ST5&+&\\
 		SRR1014709 & \textit{S. aureus} USA800-NY-313 CC5-ST83&+&\\
 		SRR1014711 & \textit{S. aureus} USA100-OR-293 CC5-ST5&+&\\
 		SRR1014712 & \textit{S. aureus} USA100-NY-76 CC5-ST5&+&\\
 		SRR1014713 & \textit{S. aureus} USA100-NRS382 CC5-ST5&+&\\
 		SRR1014714 & \textit{S. aureus} USA100-NY-54 CC5-ST105&+&\\
 		SRR1014715 & \textit{S. aureus} USA100-CA-126 CC5-ST5&+&\\
 		SRR1014716 & \textit{S. aureus} USA100-CA-248 CC5-ST5&+&\\
 		SRR1014717 & \textit{S. aureus} USA1000-CA-629 CC59-ST87&+&\\
 		SRR1014720 & \textit{S. aureus} USA200-NRS383 CC30-ST346&+&\\
 		SRR4019421 & \textit{S. aureus} TX-AML1601921&+&\\
 		SRR1014722 & \textit{S. aureus} USA600-BAA1754 CC45-ST45&+&\\
 		SRR1014724 & \textit{S. aureus} USA600-CA-347 CC45-ST45&+&\\  
 		\hline\hline
 	\end{tabular}
 	\label{table:nonlin} 
 \end{table}
 
 \section{Results}
 \subsection{SNP analysis in methicillin-resistant \textit{S.aureus}}
 The whole genome sequencing provides appropriate depth and breadth of sequence coverage across all isolates to conduct robust SNP analysis. SNP histogram analysis demonstrates the number of SNPs in sliding windows between two genomes. The distribution of SNPs along the sliding window can be considered as the genome fingerprints and can be used for genotyping. 
 
 From the SNP variant profiles of MRSA strains, we find that there are several major high-density SNPs region in the MRSA strains (Fig.1 (a) and (b)). MRSA USA300, isolated in September 2000, and has been implicated in epidemiologically unassociated outbreaks of skin and soft tissue infections in healthy individuals. The SNP histogram of MRSA USA300 is shown in Fig. 1(b). There are seven high variant regions that have SNPs at least ten a sliding window (Table 2). The four major and large high SNP density regions in the SNP histogram are mapped to four large mobile genetic elements. These elements encode resistance and virulence determinants that could enhance fitness and pathogenicity \textit{S. aureus} USA300 ~\citep{diep2006complete}. The first genetic element is Sa2usa, located at 1462625–1506875 (reference genome position, 44kb), encoding Panton-Valentine leucocidin (lukS-PV and lukF-PV genes), which is a pore-forming toxin that induces polymorphonuclear cell death by necrosis or apoptosis. The next high variant region is at 1832875-1859875(27kb), which contains two pathogenicity elements, the serine proteases A and B (SplA and SplB) ~\citep{paharik2016spl}, and island SaPIn3 encodes the leukotoxin (LukD). Panton-Valentine leucocidin is epidemiologically strongly associated with invasive disease and virulence of the USA300. The genetic element is vSA$\beta$, located at 1924875-1962375(37kb), carrying several gene clusters, for example, bacteriocin (bsa) gene, which might also contribute to pathogenesis ~\citep{baba2002genome}. The third genetic element is prophage Sa3usa, located at 2034875-2085375(50kb), encoding staphylokinase (sac gene) and a chemotaxis inhibitory protein (chp gene). Staphylokinase, fibrin-specific blood-clot dissolving enzyme, is a plasminogen activator and can enhance bacterial spreading by its blood-clot dissolving activity. The chemotaxis inhibitory protein inhibits C5$\alpha$-dependent recruitment of neutrophils. In addition to these three long high SNP regions, three short regions are identified. The first short region of high SNP pitch is located at 550375-554875, encoding fibrinogen-binding protein SdrD. The SdrD protein contributes to the pathogenicity by its capacity to bind the extracellular matrix and the host cells~\citep{josse2017staphylococcal}. The virulence characteristics of these major high SNP regions of MRSA USA300 identified by the histogram analysis agrees with experimental evidence. The highest SNP pitch is located at 405625-407375, and the other short high SNP region is located at 289625-294125. The encoded proteins in these two regions are hypothetical proteins. We postulate that the hypothetical proteins may also play important roles in the pathogenicity and virulence of the bacteria. 
 
 Our method demonstrates that high SNP density for all these critical regions correlates the genes that encode proteins for high pathogenicity. These high density SNP regions are mostly from phage integrations. We may consider these phage fragment distribution as the fingerprint of bacterial genome architecture. Phages are important vehicles for horizontal gene exchange between different bacterial species and account for a good share of the strain-to-strain differences within the same bacterial species. Therefore, this result indicates that MRSA strains might have a unique SNP genotype profile, which can be exploitable for investigating the pathogenicity and epidemiology of the MRSA strains. 
 
 \begin{figure}[tbp]
 	\centering
 	\subfloat[]{\includegraphics[width=6.0in]{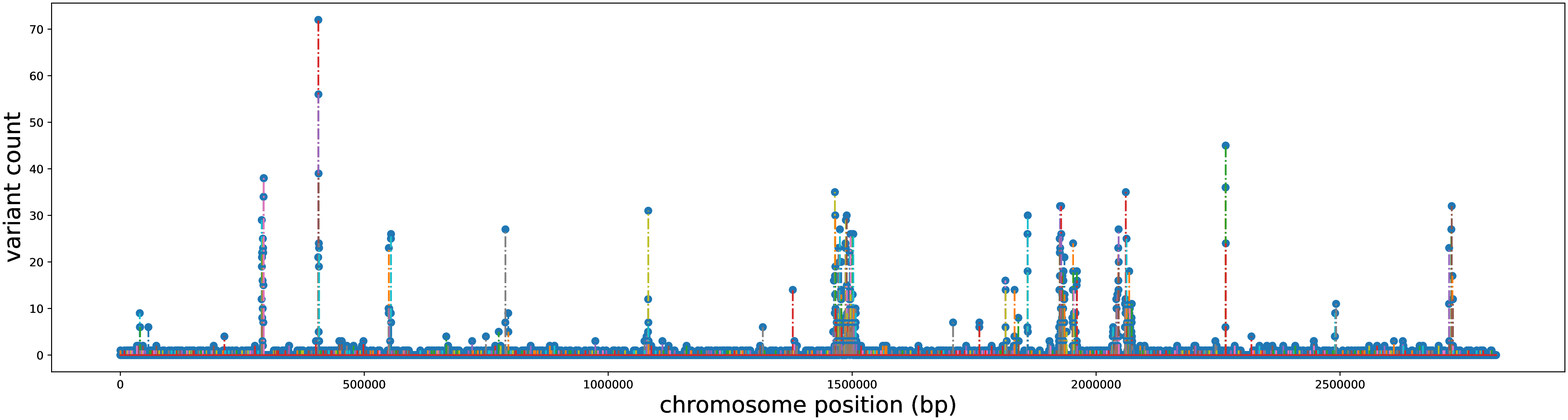}}\quad
 	\subfloat[]{\includegraphics[width=6.0in]{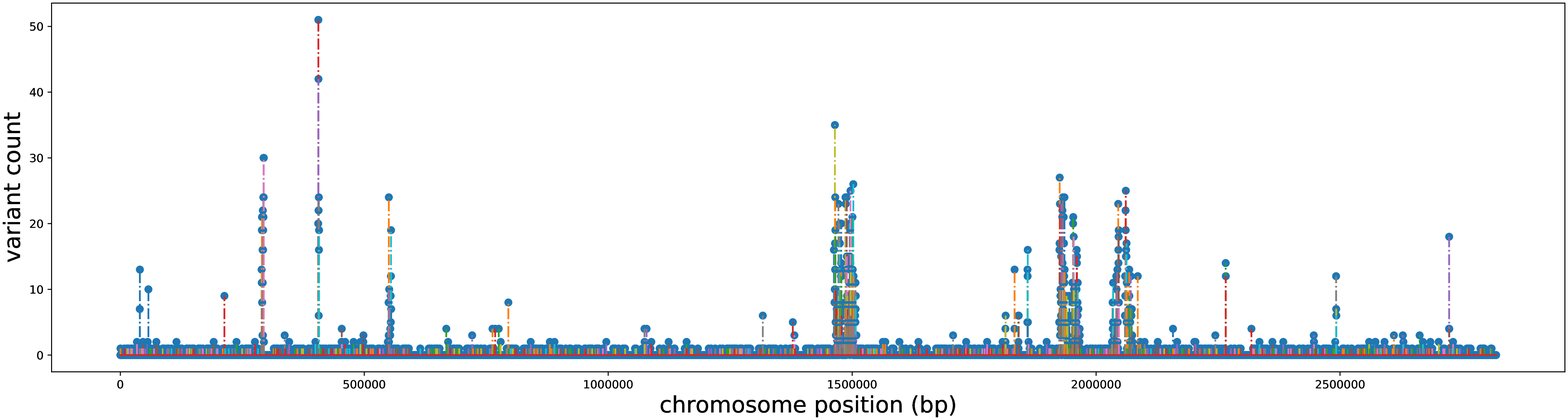}} 
 	\caption{The histograms of SNPs in whole genome sequencing of methicillin-resistant \textit{S. aureus}. The SNPs were acquired by mapping the sequencing reads to the reference genome \textit{S. aureus} NCTC8325. (a) \textit{S. aureus} TX-AML160192. (b) \textit{S. aureus} USA300 22862 R1. The histograms were calculated by the sliding window of size 250 bp.}
 \end{figure}
 
 \begin{table}[ht]
 	\caption{High SNP variant regions in \textit{S. aureus} USA300 genome ~\citep{baba2002genome,diep2006complete,josse2017staphylococcal}.}
 	\centering 
 	\begin{tabular}{l*{5}{l}r}
 		\hline\hline
 		coordinate & length & protein & gene & virulence\\
 		\hline
 		289625-294125   & 4500 & hypothetical protein & hypothetical CDS & NA \\
 		405625-407375   & 1750 & hypothetical protein & hypothetical CDS & NA \\
 		550375-554875   & 4500 & fibrinogen-binding protein & SdrD & Y \\
 		1462625-1506875 &44250 & Panton-Valentine leucocidin & lukS,lukF & Y \\
 		1832875-1859875 &27000 & leukotoxin & lukD & Y&Y\\
 		1924875-1962375 &37500 & vSA$\beta$, phage proteins & bsa & Y\\
 		2034875-2085375 &50500 & staphylokinase, chemotaxis inhibtor & sac,chp& Y \\
 		\hline\hline
 	\end{tabular}
 	\label{table:nonlin} 
 \end{table}
 
 We also analyze the SNPs that were called on different reference genomes. The next-generation sequencing reads in \textit{S. aureus} PtA02-T1 were called for variants by aligning to wild-type \textit{S. aureus} NCTC8325 genome and MRSA \textit{S. aureus} ST252 genome (Fig.2). The SNP variant profile of \textit{S. aureus} PtA02-T1 contains the same pattern as found in \textit{S. aureus} TX-AML160192 and \textit{S. aureus} ST398 (Fig. 2(a)). Two MRSA strains may have numerous random SNP variants occurred (Fig. 2(b)). These random SNP variants reflect the evolutionary dynamics of these strains.
 
 By using different reference genomes of pathogenic bacteria in SNP profile analysis, the generated SNP profiles can be used as fingerprints to potentially reveal the presence of all microorganisms in a patient sample. Since each microbial pathogen may be identified by its unique genomic fingerprint, the SNP genotyping method may allow the diagnosis of infections without prior knowledge of disease cause.
 
 \begin{figure}[tbp]
 	\centering
 	\subfloat[]{\includegraphics[width=6.0in]{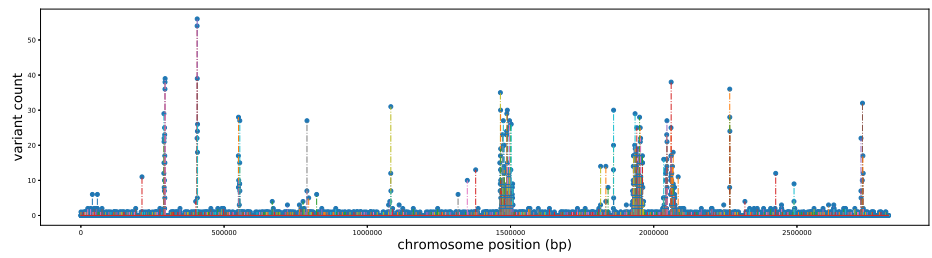}}\quad
 	\subfloat[]{\includegraphics[width=6.0in]{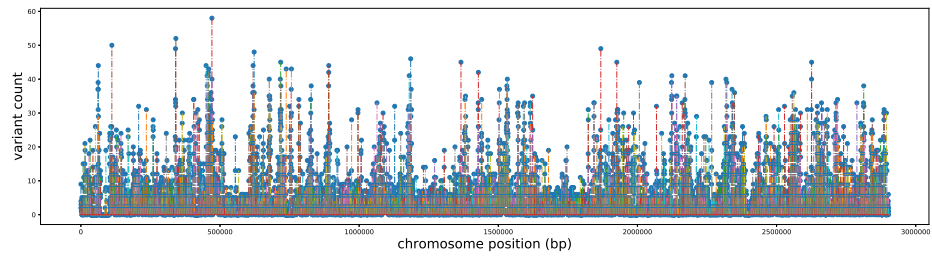}}
 	\caption{The histograms of SNPs in whole-genome sequencing of methicillin-resistant \textit{S. aureus}. (a) The SNPs of \textit{S. aureus} PtA02-T1 were obtained by aligning the sequencing reads to the reference genome \textit{S. aureus} NCTC8325. (b) The SNPs of \textit{S. aureus} PtA02-T1 were obtained by aligning the sequencing reads to the reference genome \textit{S. aureus} MRSA 252. The histograms were calculated by sliding windows of size 250 bp.}
 \end{figure}
 
 \subsection{Phylogeny of methicillin-resistant \textit{S. aureus} by rare SNP genomic regions}
 We may infer the evolution and transmission patterns of pathogenic bacteria from the whole genome SNP analysis. To this end, we measure the Jaccard distances of SNP variants of the genomes and construct the phylogenetic trees from the distance matrices.
 
 The choice of whole genome SNPs or specific region SNPs may provide different insights into the genome evolution. When estimating the evolutionary relationships among microbes using DNA sequences, the recombinations do not fit a bifurcating model. The tree representation may fail to accurately portray a reasonable genealogy. Therefore, the evolutionary inference by phylogenetic analysis is commonly performed on stable genomic regions or genes that have very low variants ~\citep{lei2018genetic}. These regions or genes have few recurrent mutations at the same sites ~\citep{huang2016new,huang2012primate}. For example, phylogenetic construction in MRSA could rely on 16S rRNA, all ribosomal-protein-coding genes, or multilocus sequence typing (MLST) housekeeping genes. These gene regions are universally present and rarely subject to high mutations, gene duplications, or lateral gene transfer.
 
 We first excluded the four long high SNPs regions, which are recombinant mobile regions 4-7 (Table 2.), and used the concatenated genetically stable regions of genomes in phylogenetic analysis. The phylogenetic tree of \textit{S. aureus} SNPs demonstrates the evolutionary relationship (Fig.3). The branch structures of the tree, which reflect the pair-wise Jaccard distances of SNPs in the genomes are well distinguishable. The phylogenetic analysis shows that \textit{S. aureus} USA 600, 700, and 800 are closely related and \textit{S. aureus}USA 100 and USA300 are very close. The strain ST398 is clustered with the housekeeping pta mutant strain PtA04-Ts. This result is consistent with previous findings \citep{hanssen2004local}.
 
 \begin{figure}[tbp]
 	\centering
 	{\includegraphics[width=6.0in]{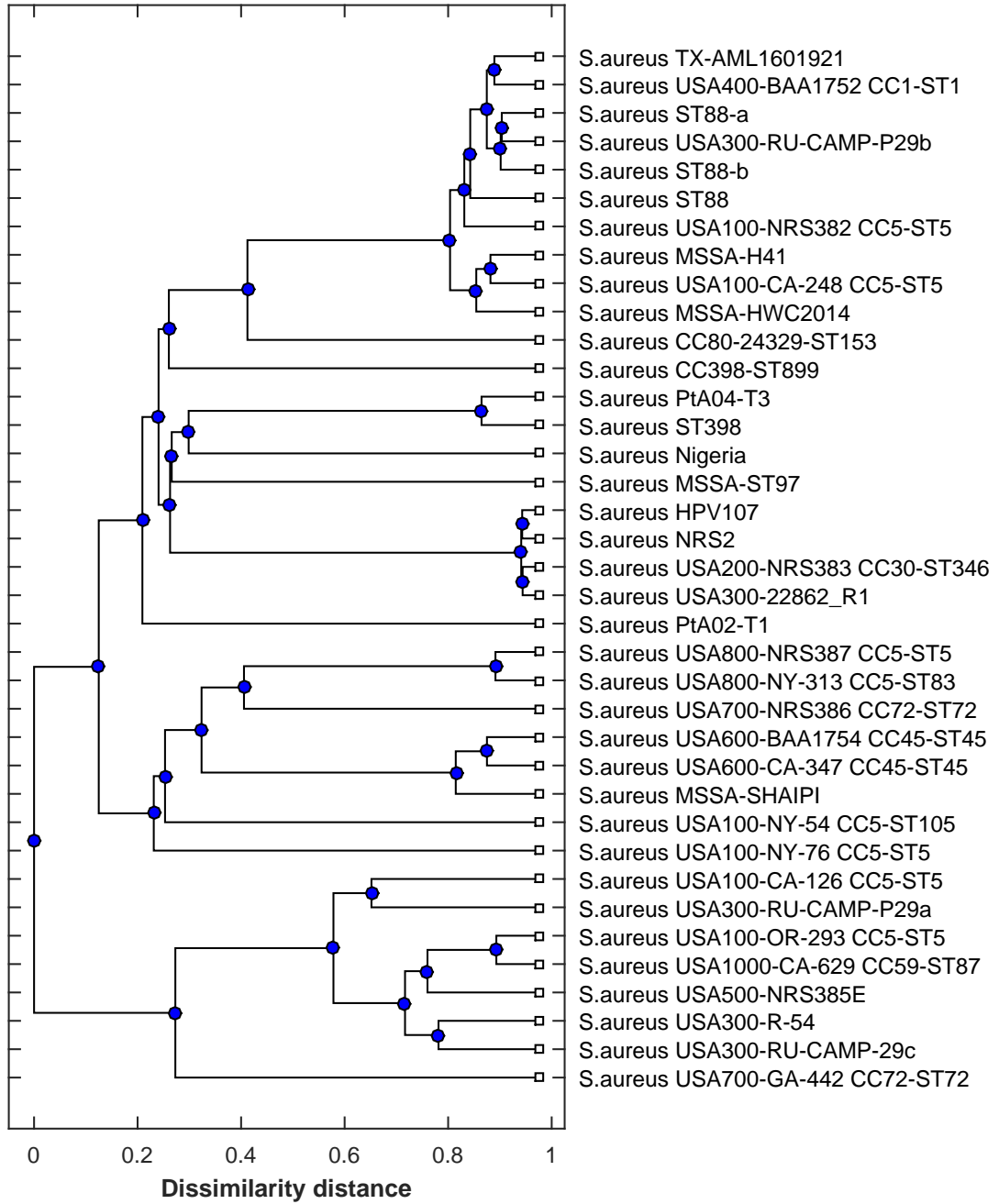}}
 	\caption{Phylogenetic analysis of MRSA and MSSA strains by conserved genomic regions. The phylogenetic tree was constructed from the pair-wise Jaccard distances of SNP profiles of the genomes. The SNP profiles were called for variants of the whole genome sequencing reads on the reference genome \textit{S. aureus} NCTC8325.}
 \end{figure} 
 
 In evolution analysis, high SNP regions usually do not follow the infinite site assumption and may cause uncertainty for building phylogenetic trees ~\citep{teske2018mitochondrial,kern2018neutral}. Genomic regions with rich SNPs have high mutation rates and are more likely to undergo recurrent mutations. When using high SNP regions or rare SNP regions, the resulting phylogenetic tree may have different topologies and only the rare SNP regions may better represent the underlying phylogeny. In bacterial genomes, the genes in high variant regions often encode proteins that interact with the host or external environment due to strong evolution selection, therefore, the high SNP regions are associated with the pathogenicity and virulence of bacteria. Despite this uncertainty in the phylogenetic analysis using high SNP variants, the phylogenetic tree reflects the distribution and transmission of the highly mutated mobile elements. We concatenate the three high variant regions of MRSA (Table 2.), and determine the pair-wise Jaccard distances of SNPs in the joined regions to construct the phylogenetic tree of \textit{S. aureus} (Fig.4). 
 
 Comparing Fig.3 and Fig.4, most of the evolution patterns are similar using either high conversed or high variant regions. For example, the strains HPV107, NRS2, and USA300-22862-R1, and USA00-NRS383 CC300-ST346 are clustered as one clad in both Fig.3 and Fig.4. PtA04-T3 and ST398 are closely related. Our result supports the advocate that diversified gene groups shall be considered as intra-species markers, as alternatives to housekeeping genes, for the phylogenetic analysis of \textit{S. aureus} ~\citep{cooper2006phylogeny}. The similar phylogenetic tree structures also suggest that the bacterial genomes and the integrated high variant phages undergo co-evolutions. This finding is in agreement with previous studies ~\citep{brussow2004phages}.
 
 \begin{figure}[tbp]
 	\centering
 	{\includegraphics[width=6.0in]{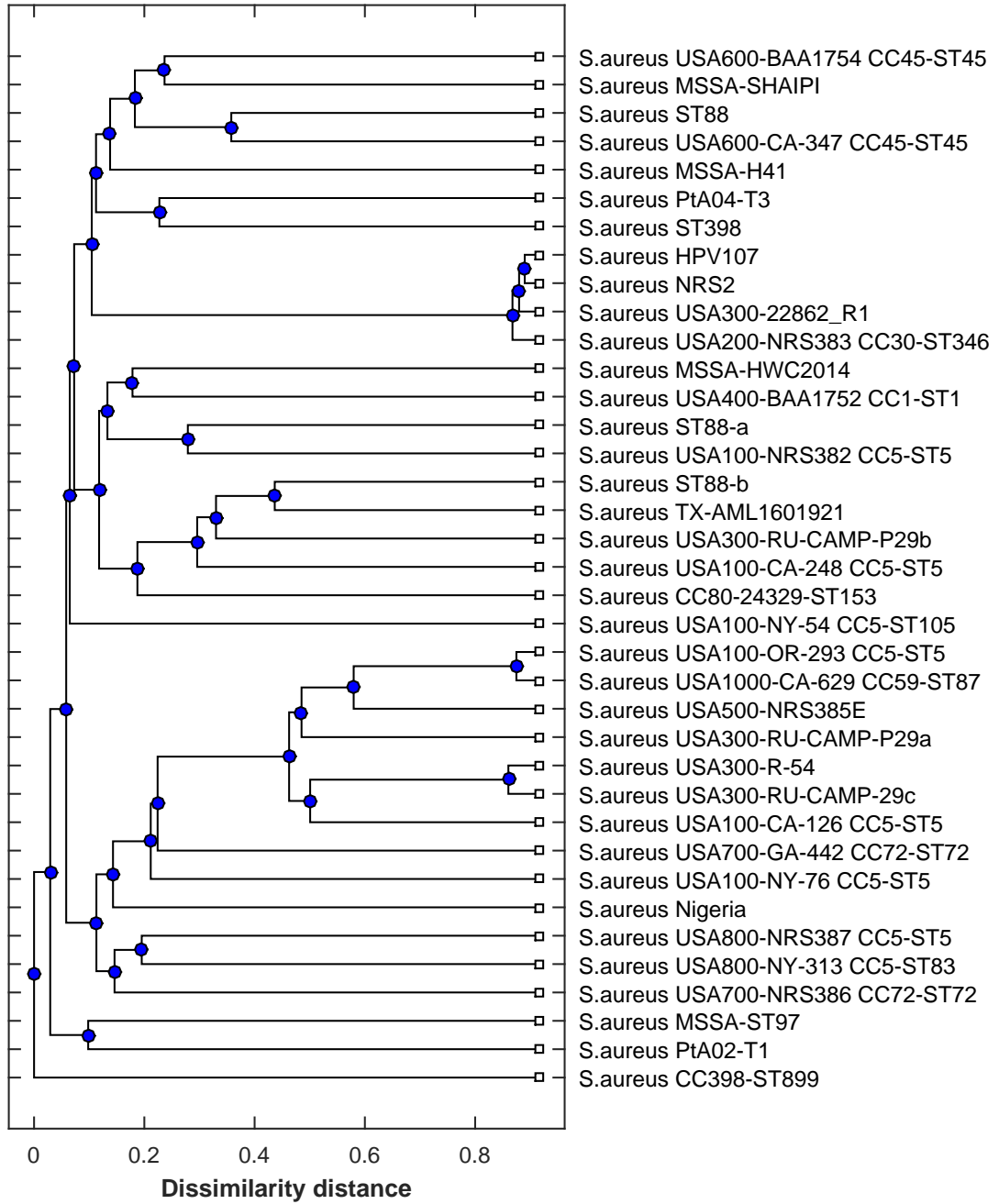}}
 	\caption{Phylogenetic analysis of MRSA and MSSA strains by high variant regions. The phylogenetic tree was constructed from the pair-wise Jaccard distances of SNP profiles of the genomes. The SNP profiles were called for variants of the whole genome sequencing reads on the reference genome \textit{S. aureus} NCTC8325.}
 \end{figure}  
 
 Despite some level of similarity in the phylogenetic trees inferred from high or rare SNP regions, there are some major differences in the phylogenies. The important difference between two SNP trees is that the most basal lineage for the rare SNP tree, USA700-GA-442-CC72-ST72, is different from the high SNP tree, CC398-ST899 (Fig.4 and Fig.5). The basal lineage is critical to identify the origin of the bacterium. Previous study shows that \textit{S. aureus} CC398 clade has a substantial heterogeneity and virulence characteristics in comparison to classical human-adapted CA- and HA-MRSA clonal lineages ~\citep{ballhausen2014mrsa}. The phylogenetic analysis agrees with this previous study. Many members of the basal lineages are important pathogens that have high repeat and transposable elements ~\citep{ma2009genomic}. Therefore, the basal lineage is an important position for the phylogenetic tree of bacteria. The other major difference between high SNP and rare SNP tree is that terminal branches of the rare SNP tree are significantly shorter. In addition, when the terminal branches are short in the high SNP tree, such as the clade containing HPV107, the same clade topology is also found in the rare SNP tree. But when the terminal branches are long in the high SNP tree, such as the clade containing ST97 and PtA02-T1, a different clade topology would be found in the rare SNP tree where ST97 now groups with others such as ST398 and PtA04-T3. This phenomenon is consistent with the expectation based on the idea of maximum genetic distance. Short branch length in the high SNP region tree in a clade means short time of divergence among lineages of the clade and because of the short time, the less conserved DNA region is still at the linear phase of accumulating genetic distances and hence are expected to be equally informative as the conserved region in phylogenetic inferences. Therefore, both high SNP and rare SNP regions can produce identical tree topology. In contrast, for clades with very long branch length in the high SNP tree, the distance between two lineages is likely to be at the maximum saturation level and hence not informative anymore to separation time between lineages. In this case, only the rare SNP tree could be informative. Therefore, the rare SNP tree is better to depict the evolutionary history of genomes.
 
 \subsection{Phylogeny of methicillin-resistant \textit{S. aureus} by whole genomes}
 The phylogeny using whole genomes, called phylogenomics, is presented (Fig.5). By comparing entire genomes among many species, phylogenomics demonstrates the global genomic relationship and possibly identifies laterally transferred genes ~\citep{delsuc2005phylogenomics}. The strains are clustered as three major lineages, USA100-USA300, USA600-USA700, and ST88. Both USA300 and ST88 are community-acquired MRSA ~\citep{kpeli2017genomic} and clustered as a clade. The phylogenetic analysis shows that USA600-CA347 has a similar SNP pattern as MSSA-SHAIPI, and USA100-100-CA has a comparable SNP pattern as MSSA-HWC2914. This result may infer these two Community Acquired-MRSA strains are potentially derived from MSSA strain. This result indicates that the genetic relatedness among isolates can be determined using whole-genome SNP data. The phylogenetic analysis based on SNP genotyping may monitor and construct the transmission pathways of MRSA. 
 
 It should be noted that the phylogenetic analysis using whole genome sequencing may not illustrate the exact sequence elements in antibiotics resistances and virulence of bacteria. The factors of high virulence of a bacteria are complex and yet no certain solution to the antibiotics resistance of MRSA has been found. Our study of SNP accumulation may provide a clue for potential high virulence.   
 
 \begin{figure}[tbp]
 	\centering
 	{\includegraphics[width=6.0in]{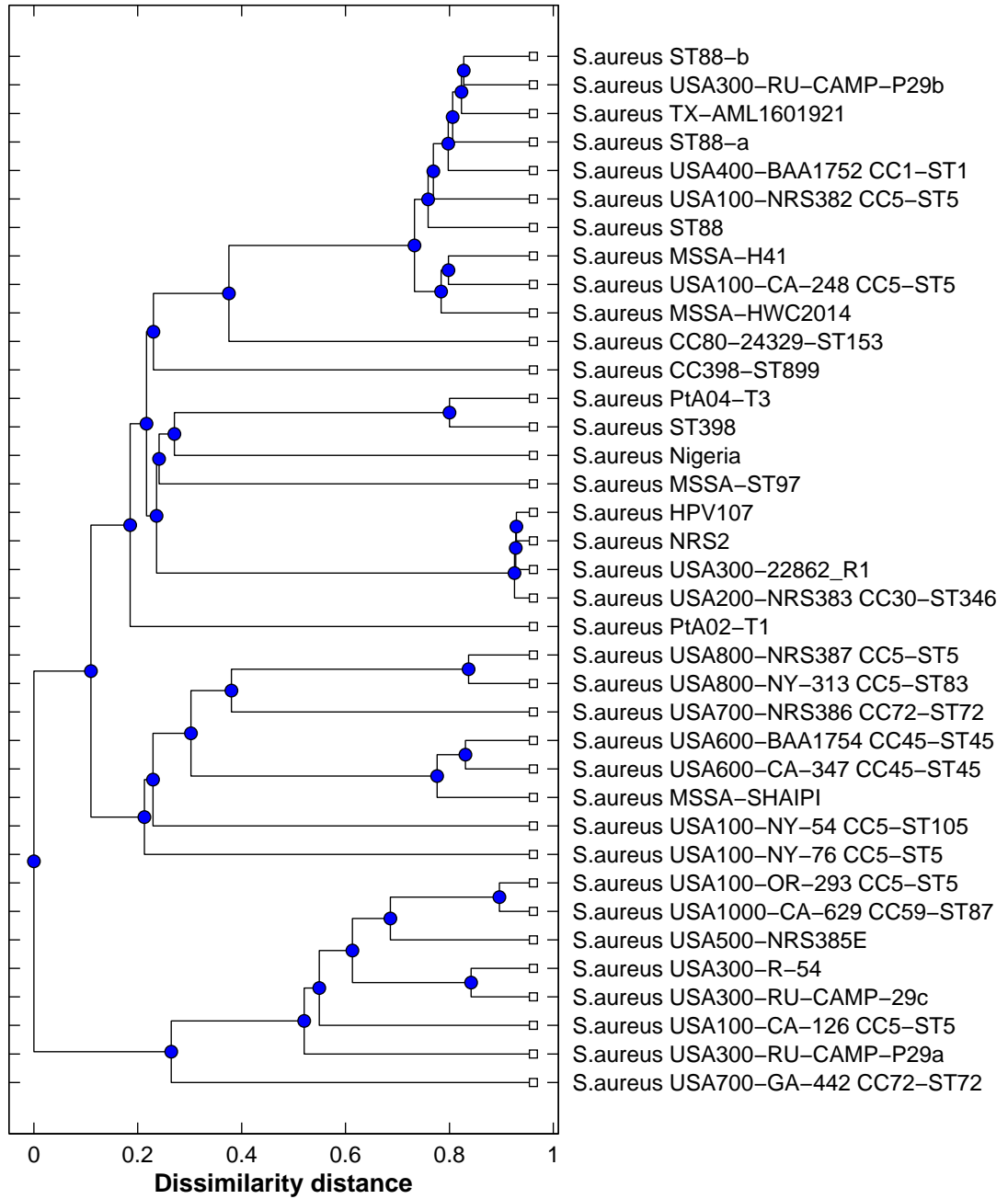}}
 	\caption{Phylogenetic analysis of MRSA and MSSA strains by whole genomes. The phylogenetic tree was constructed from the pair-wise Jaccard distances of SNP profiles of the genomes. The SNP profiles were called for variants of the whole genome sequencing reads on the reference genome \textit{S. aureus} NCTC8325.}
 \end{figure}  
 
 \subsection{SNP genotyping reveals structural variants in MSSA \textit{S. aureus} genomes}
 Most serious \textit{S. aureus} infections in the community and some in hospitals are caused by methicillin-susceptible isolates (MSSA), yet there are few attempts to identify the structural variants including deletions and insertions of hypervirulent MSSA in serious disease. The SNP genotyping approach may reveal the deletions of a genome when the SNP profile is compared with the reference genome. 
 
 We selected three MSSA whole genome sequencing reads to investigate the structural variants in the MSSA genomes. From the SNP genotyping analysis, a large deletion region at positions from 1462750 to 1493500 is inferred by long zero SNPs in the histogram graph. A large deletion region of 30.75kb is found in \textit{S. aureus} MSSA-H41 (Fig.6(a)) and MSSA-HWC2014 (Fig.6(b)). The same deletions are observed in other MSSA strains including \textit{S. aureus} MSSA-ST97. Further examination suggests that the deletion region contains numerous phage genes, which encode and disseminate potent staphylococcal virulence factors ~\citep{deghorain2012staphylococci}. For example, the deletion region encodes the Panton-Valentine leukocidin(PVL)(locus tag=SAOUHSC\_01553) to form pores into leukocytes and cause necrotic infections. We may infer that when some critical phage regions are deleted in MSSA strains, the strains become methicillin-susceptible and then less virulent. On the other hand, we may also postulate that when a strain acquire some special phage segments, the phage segments possibly account for the increased virulence and methicillin resistance.
 
 \begin{figure}[tbp]
 	\centering
 	\subfloat[]{\includegraphics[width=6.0in]{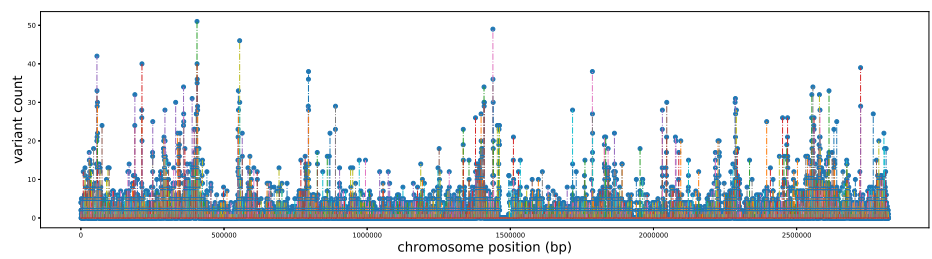}}\quad
 	\subfloat[]{\includegraphics[width=6.0in]{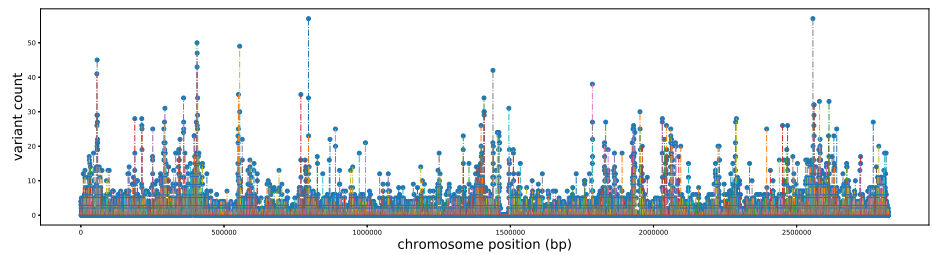}}
 	\caption{The histograms of SNPs in whole genome sequence of MSSA strains. The SNPs were obtained by mapping sequencing reads to the reference genome \textit{S. aureus} NCTC8325. (a) \textit{S. aureus} MSSA-H41. (b) \textit{S. aureus} MSSA-HWC2014} 
 \end{figure}
 
 \section{Discussion}  
 Many infectious diseases have similar signs and symptoms, making it challenging for health-care providers to precisely identify the infectious agent. Analysis of the entire pathogen genome via whole genome sequencing could provide unprecedented resolution in discriminating even highly related lineages of bacterial pathogens and revolutionize outbreak analysis in hospitals. Nevertheless, clinicians have long been hesitant to implement whole genome sequencing in outbreak analyses due to the expensive and cumbersome nature of early sequencing platforms. Recent improvements in sequencing technologies and analysis tools have rapidly increased the output and analysis speed as well as reduced the overall costs of whole-genome sequencing. The advantage of whole genome sequencing-based SNP genotyping presented in this study is that the method solely depends on the raw sequencing reads, not assembled the whole genome. This will be a rapid process when an outbreak occurs. In addition, the SNP genotyping may give an excellent resolution on gene mutations and structural variants. 
 
 From SNP analysis of MRSA strains, we may confirm that some increased virulence in community-acquired (CA) MRSA is attributed by many CA organisms harboring genes(lukS-PV and lukF-PV). These genes encode the subunits of PVL. An understanding of community acquisition and transmission should lead to more enlightened interventions aimed at treatment and prevention of CA-MRSA infection. CA-MRSA has emerged as a significant public health problem in the USA. We demonstrate that the whole-genome SNP genotyping represent an effective method to detect and monitor the epidemiology of MRSA. These methods hold promise in directing infection control more efficiently.
 
 The limitation of the SNP-based approach is that a reference genome closely related to the sequence sample is required for alignment. The resulting alignments constitute the basis for the variant calling and phylogenetic analysis reconstruction. To align various bacterial strains, the same reference genome should be used across diverse samples. When the reference genome is not available, which can be inferred by large missing coverages in mapping, then this approach cannot produce reliable results. 
 
 The evolutionary inference by phylogenetic tree is usually performed on low variant regions or genes ~\citep{lei2018genetic}. These regions or genes undergo few recurrent mutations at the same sites ~\citep{huang2016new,huang2012primate}. The regions with rich SNPs may have high mutation rates and are more likely to undergo recurrent mutations. Therefore, it is expected that high SNP regions may not follow the infinite site assumption and may cause uncertainness for building phylogenetic trees ~\citep{teske2018mitochondrial,kern2018neutral}. Despite some similar topologies in high SNP and SNP trees, the rare SNP may provide more meaningful evolutionary history of bacteria.
 
 \section{Conclusion}
 The whole genome sequencing-based SNP genotyping method proposed in this study may capture the fingerprint character of a genome and perform the phylogenetic analysis. The phylogenetic analysis using whole genome sequencing is superior to conventional epidemiologic methods for monitoring and constructing transmission pathways of MRSA. The whole genome sequencing-based SNP genotyping approach, together with the antimicrobial susceptibility testing, will provide deep insights into the evolutionary and transmission dynamics of \textit{S. aureus}. Therefore, the whole-genome sequencing SNP analysis represents an ideal method to monitor the pathogen infection and spreads.
 
\section*{Acknowledgement}
This research was supported by National Natural Sciences Foundation of China (31271408 to S.S.-T.Yau) and Tsinghua University start-up research fund (to S.S.-T.Yau).
We are grateful to Dr.Donald Morrison at University of Illinois at Chicago and Prof.Jiasong Wang at Nanjing University for insightful discussion and valuable comments. We thank Melissa Yin for collecting genome data. We appreciate two anonymous reviewers for constructive review so that the research was greatly improved.

\newpage

\bibliographystyle{elsarticle-harv}
\bibliography{../References/myRefs}



\end{document}